\documentclass[twocolumn,superscriptaddress,aps,prl]{revtex4}
\usepackage{amsfonts}
\usepackage{amsmath}
\usepackage{amssymb}
\usepackage{graphicx}
\usepackage{color}
\usepackage{ulem}
\usepackage{dcolumn}
\usepackage{bm}
\usepackage[colorlinks,urlcolor=cyan,citecolor=blue,linkcolor=magenta]{hyperref}

\begin{document}

\title{Density Dependent Spin-Orbit Coupling in Degenerate Quantum Gases}
\author{Peng Xu}
\affiliation{Institute for Advanced Study, Tsinghua University, Beijing, 100084, China}
\author{Tianshu Deng}
\affiliation{Institute for Advanced Study, Tsinghua University, Beijing, 100084, China}
\author{Wei Zheng}
\email{zw8796@ustc.edu.cn}
\affiliation{Hefei National Laboratory for Physical Sciences at the Microscale and
Department of Modern Physics, University of Science and Technology of China,
Hefei 230026, China}
\affiliation{CAS Center for Excellence in Quantum Information and Quantum Physics,
University of Science and Technology of China, Hefei 230026, China}
\author{Hui Zhai}
\email{huizhai.physics@gmail.com}
\affiliation{Institute for Advanced Study, Tsinghua University, Beijing, 100084, China}
\date{\today }

\begin{abstract}
In this letter we propose a method to realize a kind of spin-orbit coupling
in ultracold Bose and Fermi gases whose format and strength depend on
density of atoms. Our method combines two-photon Raman transition and
periodical modulation of spin-dependent interaction, which give rise to the
direct Raman process and the interaction assisted Raman process, and the
latter depends on density of atoms. These two processes have opposite
effects in term of spin-momentum locking and compete with each other. As the
interaction modulation increases, the system undergoes a crossover from the
direct Raman process dominated regime to the interaction assisted Raman
process dominated regime. For this crossover, we show that for bosons, both
the condensate momentum and the chirality of condensate wave function change
sign, and for fermions, the Fermi surface distortion is inverted. We
highlight that there exists an emergent spatial reflection symmetry in the
crossover regime, which can manifest itself universally in both Bose and
Fermi gases. Our method paves a way to novel phenomena in a non-abelian
gauge field with intrinsic dynamics.
\end{abstract}

\maketitle

Spin-orbit (SO) coupling is an unambiguous effect in electron gases in
quantum materials. For electrons, the SO coupling is essentially a
relativistic effect of charged particles, and for a given dispersion in
solid, both the form and the strength of the SO coupling are fixed~\cite%
{SOC_In_Solids@Kittel.1963,SOC_In_Solids@Winkler.2003}. Although the
ultracold atoms are neural, SO coupling effect can now be simulated for
ultracold atoms by ultilizing the atom-light interaction~\cite%
{SOC_Rev@Zhai.2012,SOC_ReV@Spielman.2014,SOC_Rev@Zhai.2015}. In the simplest
and most widely used setting, a pair of Raman lasers are applied to
ultracold atoms~\cite%
{SOC_BEC@Spielman.2011,SOC_fermion@Zhang_Jin.2012,SOC_fermion@Zwierlein.2012}%
. These pair of lasers can flip the spin from down to up, accompanied by a
momentum transfer to the right, and simultaneously, can flip the spin from
up to down, accompanied by a momentum transfer to the left. In this way, the
spin and momentum are locked which realizes the SO coupling effect. Studying
SO coupling in ultracold atomic gases can significantly enrich our
understanding of this effect . For instance, the effect of SO coupling in a
Bose gas~\cite{
SOC_BEC@Galitski.2008,SOC_BEC@Zhai.2010,SOC_BEC@Jian_CM.2011,SOC_BEC@Ho_TL.2011, SOC_BEC@Wu_CJ.2011,SOC_BEC@You_Li.2011,SOC_BEC@Santos.2011,SOC_BEC@Chen_Shuai.2012, SOC_BEC@Baym_01.2012,SOC_BEC@Baym_02.2012,SOC_BEC@Li_Yun_01.2012,SOC_BEC@Martone.2012, SOC_BEC@Zhang_Peng.2013,SOC_BEC@Clark.2013,SOC_BEC@Demler.2013, SOC_BEC@Li_Yun.2013,SOC_BEC@Zhou_Qi.2013,SOC_BEC_finite-T@Chen_Shuai.2014, SOC_BEC@Chen_Shuai.2015,SOC_BEC@Chen_Shuai.2016}
and its interplay with the BEC-BCS crossover~\cite{
SOC_FG@Shenoy.2011,SOC_FG@Zhai.2011,SOC_FG@Zhang.2011,SOC_FG@Hu_Hui.2011,
SOC_FG@Zhang_Peng.2012,SOC_FG@Zhang_Peng.2013,SOC_FG@Shenoy.2013,SOC_FG@Melo_01.2012, SOC_FG@Melo_02.2012,SOC_FG@Zhang_Jing.2014,SOC_FG@Zhang_Jing.2016}
are both novel effects revealed by ultracold atomic systems, which have no
counterpart in electronic system studied before.

Another unique aspect of SO coupling in ultracold atom systems is that the
coupling itself can be made dynamical. That is to say, the dynamics of atoms
in the presence of SO coupling can feedback to the coupling form or strength
itself. There are two approaches to realize such dynamical SO couplings. One
approach is to replace the classical light field with quantum photon field
strongly coupled to atoms, for instance, by using cavity field in strong
coupling regime~\cite%
{Cavity_SOC@Lev.2019,Cavity_SOC@Yi_Su.2014,Cavity_SOC@Yi_Wei.2015}. Another
approach is to make the SO coupling depending on atom field itself, for
instance, depending on the density of atoms. Here we will focus on the
second approach. The SO coupling can also be viewed as a non-abelian gauge
field. Actually, for Abelian gauge field, the $U(1)$ gauge field, manifested
as the phase of hopping in optical lattices, has been made density
dependent, either by periodically shaking the optical lattice with a
frequency resonant with interaction energy~\cite{Density_SOC@Esslinger.2019}%
, or by periodically driving both optical lattice and interaction~\cite%
{Density_SOC@Chin.2018}. This also enables recent realization of dynamical
abelian gauge field with local gauge symmetry~\cite{Z2_LGT@Bloch.2019}. In
this letter, we will show that similar method can also used to realize
dynamical SO coupling in both degenerate Bose and Fermi gases. This will be
an important step toward realizing dynamical non-abelian gauge field.

\begin{figure}[t]
\centering
\includegraphics[width=0.98\linewidth]{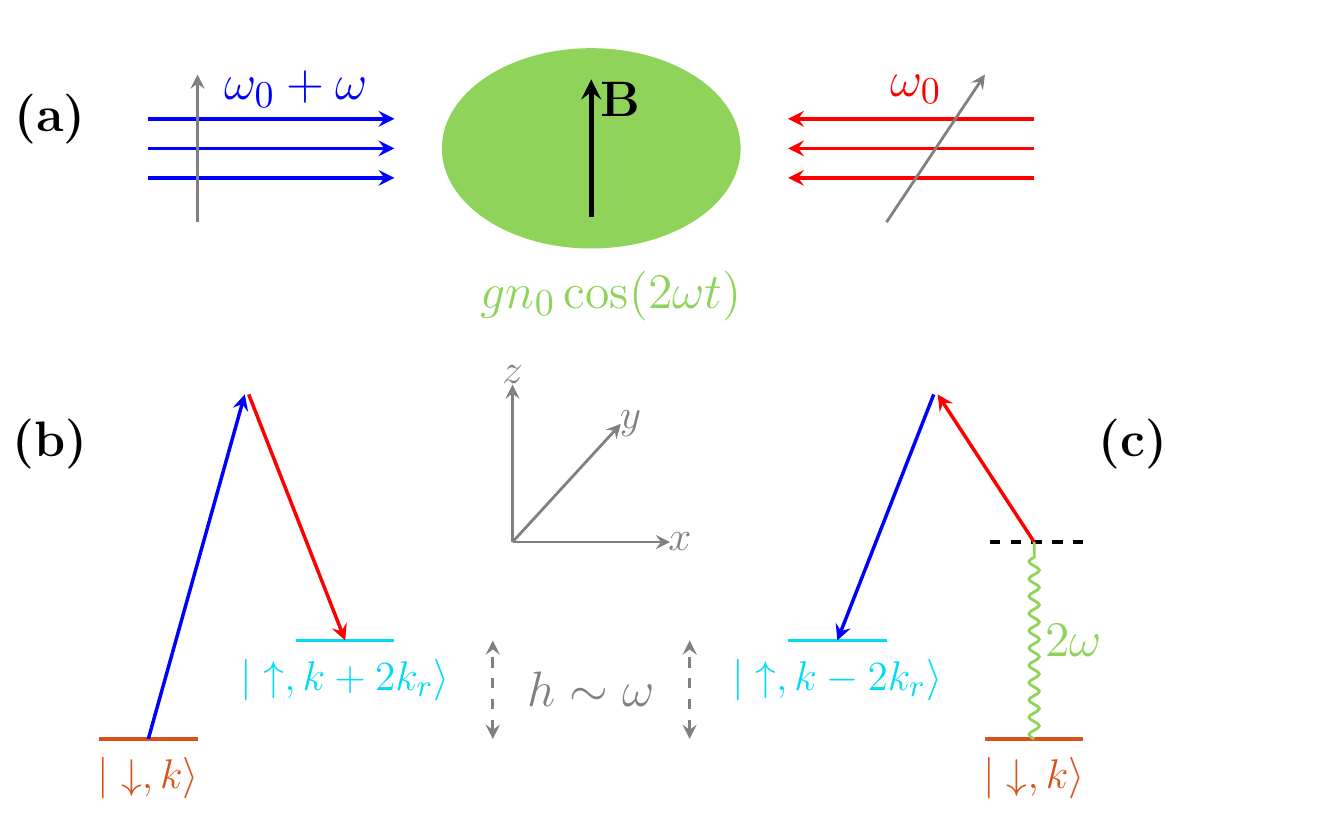}
\caption{(a) Schematic of experimental configuration. A pair of Raman lasers
with different polarization are applied to a cloud of ultracold atoms. (b)
Direct Raman transition regime. (c) Interaction assisted Raman transition
regime. For the same spin flip process, the momentum transfer are opposite
between (b) and (c). }
\label{schematic}
\end{figure}

\textit{Setting.} First, we consider the conventional configuration where a
cloud of ultracold atoms are placed in two counter propagating Raman beams,
as shown in Fig. \ref{schematic}(a). The single particle Hamiltonian is
given by
\begin{equation}
\hat{H}_{0}=-\frac{\hbar ^{2}\nabla ^{2}}{2m}+\frac{h}{2}\sigma _{z}+\hbar
\Omega \cos (2k_{\text{r}}x-\omega t)\sigma _{x}.  \label{H0}
\end{equation}%
Here $h$ is the Zeeman energy between spin-up and spin-down, $k_{\text{r}}$
is the wave length of the lasers, $\Omega $ is the strength of the Raman
process, and $\omega $ is the frequency difference between two Raman lasers.
Now we apply a unitary rotation $\hat{\mathcal{U}}_{1}=e^{-i\omega \sigma
_{z}t/2}$ to Eq. (\ref{H0}), it yields
\begin{align}
\hat{H}_{0}=& -\frac{\hbar ^{2}\nabla ^{2}}{2m}+\delta \sigma _{z}+\frac{%
\hbar \Omega }{2}(e^{i2k_{r}x}\sigma ^{+}+e^{-i2k_{r}x}\sigma ^{-})  \notag
\\
& +\frac{\hbar \Omega }{2}(e^{-i2k_{r}x+2\omega t}\sigma
^{+}+e^{i2k_{r}x-2\omega t}\sigma ^{-}).  \label{rotationH0}
\end{align}%
where $\delta =(h-\hbar \omega )/2$. Usually, considering the situation $%
h\sim \hbar \omega $, we implement the rotating wave approximation to drop
the high frequency term, that is the last term in Eq. (\ref{rotationH0}).
The retained Raman coupling term is shown in Fig. \ref{schematic}(b), where
momentum of an atom increases $2k_{\text{r}}$ when its spin is flipped from
down to up, and decreases $2k_{\text{r}}$ when it is spin is flipped from up
to down. To distinguish with another process discussed below, we refer this
process as the direct Raman coupling. With the direct Raman coupling process
only, the Hamiltonian can be equivalently written as
\begin{equation}
\hat{H}_{0}=\frac{\hbar ^{2}}{2m}(k_{x}+k_{\text{r}}\sigma _{z})^{2}+\delta
\sigma _{z}+\frac{\hbar \Omega }{2}\sigma _{x}.  \label{direct_Raman}
\end{equation}%
The physical effect of this kind of SO coupling has been discussed
extensively in the ultracold atom literatures in the past decade~\cite%
{SOC_Rev@Zhai.2012,SOC_ReV@Spielman.2014,SOC_Rev@Zhai.2015}.

\begin{figure}[t]
\centering
\includegraphics[width=0.98\linewidth]{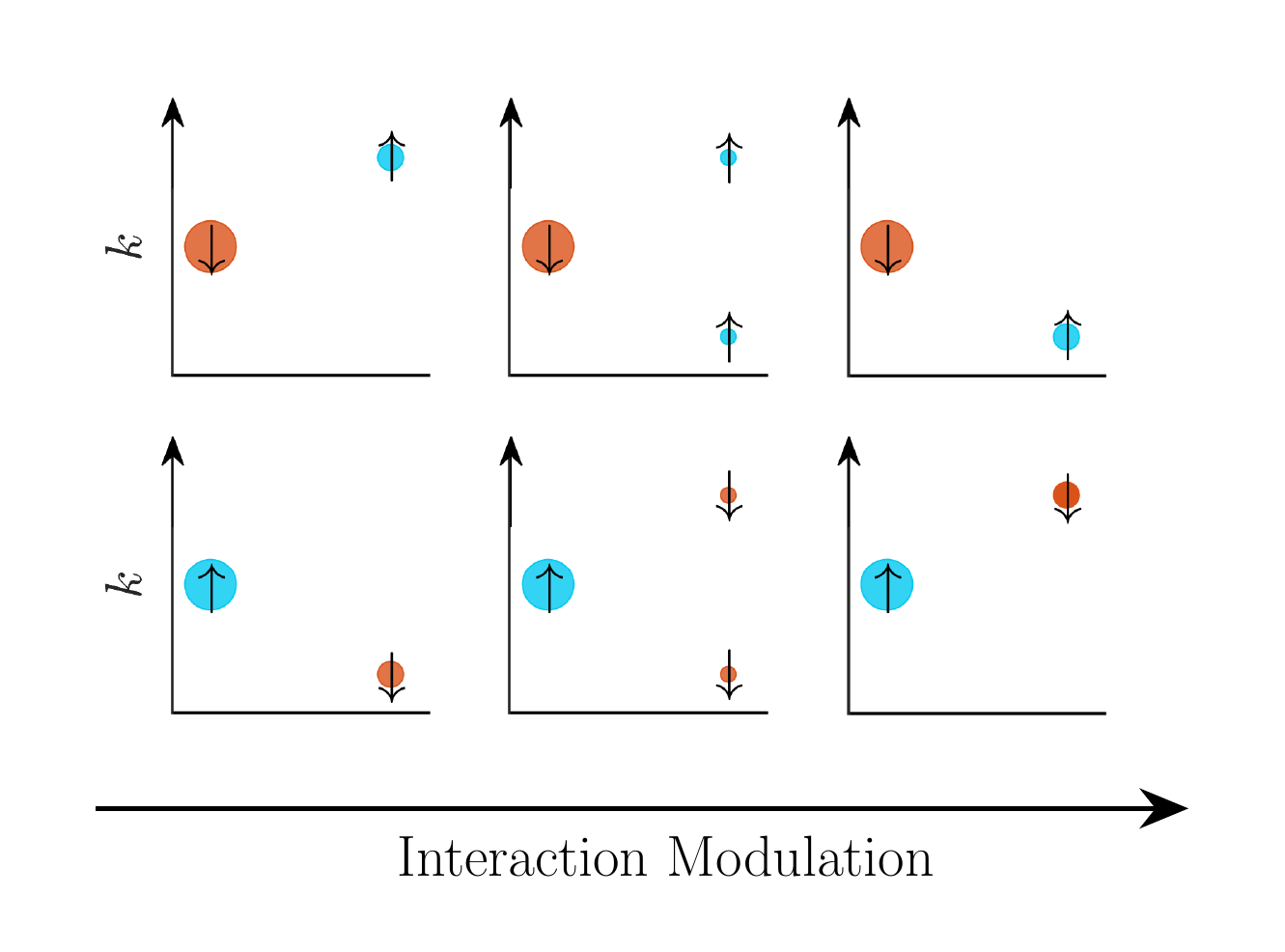}
\caption{Schematic of experimental predication for spin resolved momentum
distribution measured by the time-of-flight with the Stern-Gerlach
experiment. The first column dominated by direct Raman coupling regime, the
last column dominated by the interaction assisted Raman coupling, and the
middle column is the crossover regime where both processes are important.
For bosons, the predication is for $\protect\delta \sim 0$. For fermions,
the upper raw is for $\protect\delta\gtrsim E_\text{F}>0$ and the lower raw
is for $\protect\delta<0$ and $|\protect\delta|\gtrsim E_\text{F}$. }
\label{experiment}
\end{figure}

Let us now revisit the term ignored by the rotating wave approximation. This
term actually does opposite compared with the direct Raman process. The
momentum of an atom decreases $2k_{\text{r}}$ when its spin is flipped from
down to up, and increases $2k_{\text{r}}$ when it is spin is flipped from up
to down, as shown in Fig. \ref{schematic}(c). However, when $h\sim \hbar
\omega $, the energies between the initial and the final states of this
process differs by $\sim 2\hbar \omega $. That is also the reason why it can
be safely ignored by the rotating wave approximation. However, the situation
changes if there exists the periodic driven with frequency $2\omega $ and
coupled to spin degree of freedom, therefore, this $2\omega $ energy offset
can be compensated by this driving \cite{footnote}. Here we consider periodically modulating
spin-dependent interaction with a frequency $2\omega $. Such a technique of
time periodically modulating interaction is nowadays quite matured in
ultracold atom experiments. Thus, the combination of interaction modulation
and Raman beam gives rise to another interaction assisted Raman process, as
shown in Fig. \ref{schematic}(c). Thus, when interaction modulation is weak,
the direct process dominates. And when the interaction modulation becomes
strong, the interaction assisted Raman process dominates. In the regime
where the interaction assisted dominates, the effective Hamiltonian can be
written as
\begin{equation}
\hat{H}_{0}=\frac{\hbar ^{2}}{2m}(k_{x}-k_{r}\sigma _{z})^{2}+\delta \sigma
_{z}+\frac{\hbar \tilde{\Omega}}{2}\sigma _{x}.  \label{Interaction_Raman}
\end{equation}%
Note that $-k_{\text{r}}\sigma _{z}$ in Eq. (\ref{direct_Raman}) is now
changed to $k_{\text{r}}\sigma _{z}$ in Eq. (\ref{Interaction_Raman}), and $%
\tilde{\Omega}/2$ in Eq. (\ref{Interaction_Raman}) is the effective
interaction assisted Raman coupling strength, which depends on density.

\textit{Main Results and Experimental Predications.} Before we introduce the
details of our derivations, let us first describe the main results as shown
in Fig. \ref{experiment}, which are the experimental predications as the
interaction modulation increases.

\begin{figure}[t]
\centering
\includegraphics[width=0.98\linewidth]{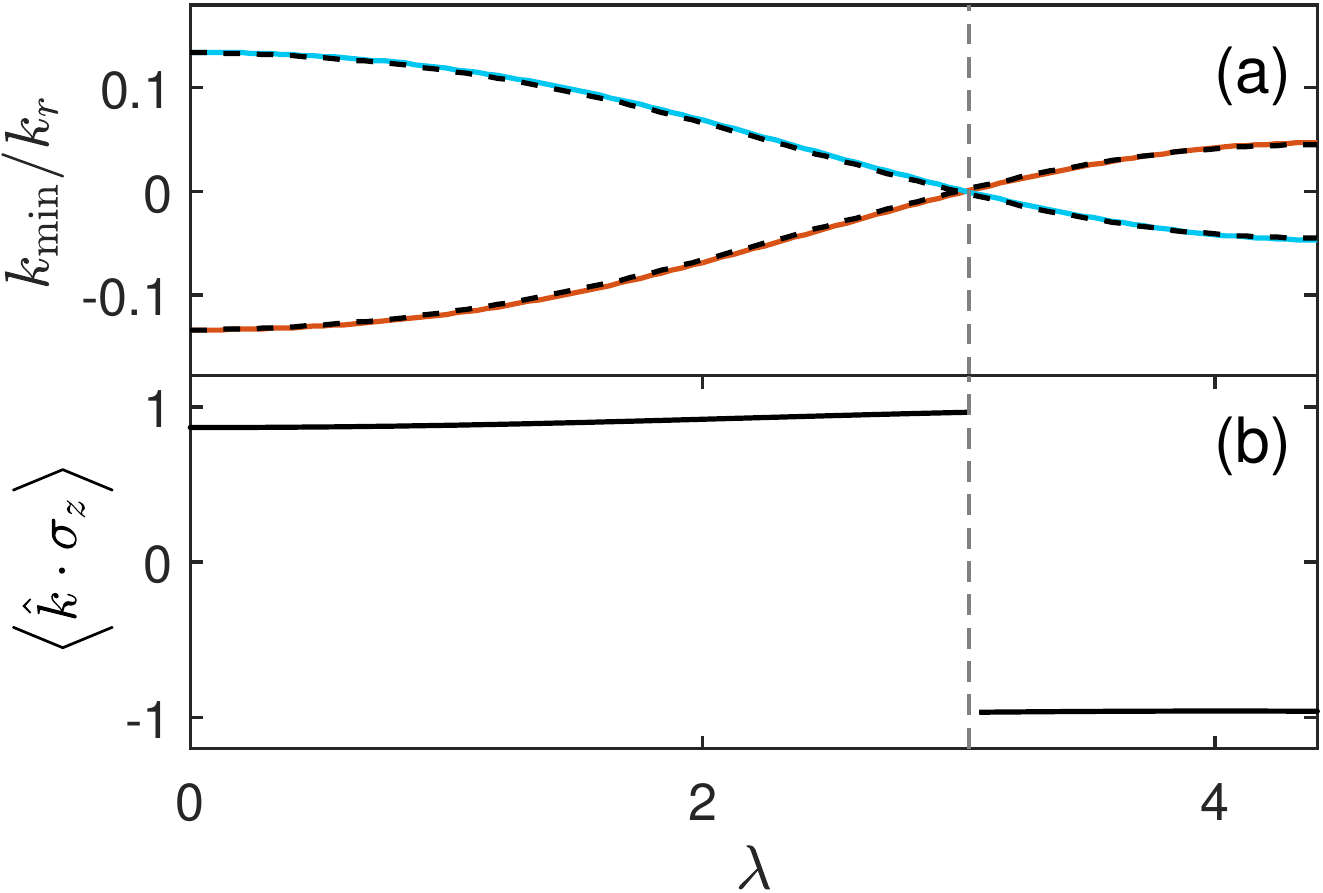}
\caption{Bose-Einstein condensate with density-dependent SO coupling. (a)
The momentum of energy minima as a function of interaction modulation
amplitude $\protect\lambda=gn_0/(\hbar\protect\omega)$. Two lines are for
two degenerate ground states with $\protect\sigma_z>0$ and $\protect\sigma%
_z<0$, respectively. (b) The chirality $\langle\hat{k}\protect\sigma%
_z\rangle $ of the ground state. The two ground states share the same value
of chirality. Here we take $\protect\delta=0$ and $\Omega=2E_r$, with $E_r =
\hbar^2 k_r^2 / (2 m)$. }
\label{Bose}
\end{figure}

For Bose condensation at zero temperature, we consider the situation $%
\delta=0$. In this case there are two degenerate ground states, where the
majority atoms, either spin-up or spin-down, condense at the momentum
minimum $k_\text{min}$ around zero. For one state, in the direct Raman
coupling regime, majority spin-down atoms are coupled to minority spin-up
atoms with positive momentum, and therefore, $k_\text{min}$ is pushed to
slightly negative value. If we define a chirality $\langle \hat{k}%
\sigma_z\rangle$ with $\hat{k}=\vec{k}/|\vec{k}|$, the chirality is
positive. In the interaction assisted Raman coupling regime, in contrast,
majority spin-down atoms are coupled to minority spin-up atoms with negative
momentum, and therefore, $k_\text{min}$ is pushed to slightly positive
value. Then the chirality is negative. In the crossover regime, both direct
and interaction assisted Raman processes are equally important, and the
minority spin-up atoms appears at both positive and negative momenta. This
is schematically shown in the first raw of Fig. \ref{experiment}. The
results from quantitative calculation is shown in Fig. \ref{Bose}, where we
show how the momentum minimum and chirality changes when the interaction
modulation increases. Especially, one can see that the chirality jumps from
positive to negative at the point where $k_\text{min}$ crosses zero. For
another degenerate state, the majority spin-up atoms are coupled to minority
spin-down atoms with negative momentum in the direct Raman coupling regime,
and are coupled to minority spin-down atoms with positive momentum in the
interaction assisted Raman coupling regime, as shown in the second raw of
Fig. \ref{experiment}. Consequently, $k_\text{min}$ changes from positive to
negative as interaction modulation increases. The behaviors of the chirality
are the same for these two states, as shown in Fig. \ref{Bose}(b).

\begin{figure}[t]
\centering
\includegraphics[width=0.98\linewidth]{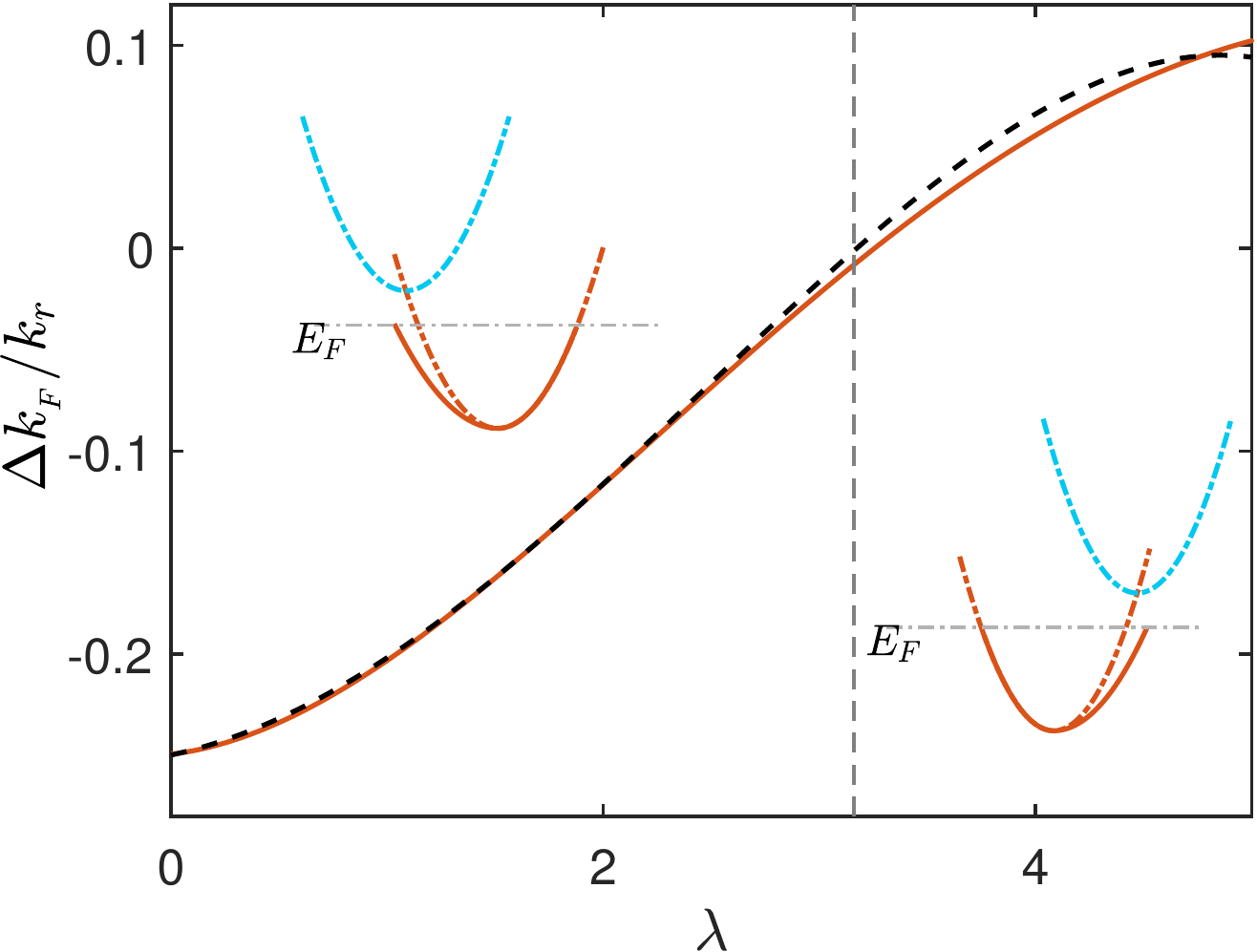}
\caption{One-dimensional degenerate Fermi gas with density-dependent SO
coupling. $\Delta k_\text{F}=k_\text{F}^{+}+k_\text{F}^{-}$, where $k^{\pm}_%
\text{F}$ is the Fermi points at positive and negative momenta. $\Delta k_%
\text{F}$ is plotted as a function of interaction modulation amplitude $%
\protect\lambda=gn_0/(\hbar\protect\omega)$. The inset schematically shows
the distortion of fermion dispersion in the direct Raman coupling regime
(left) and the density assisted Raman coupling regime (right). Here we set $%
\protect\delta=6E_r$ and $\Omega=2E_r$. }
\label{Fermi}
\end{figure}

For degenerate Fermi gas, we consider a simpler situation of one-dimension.
When $\delta \gtrsim E_\text{F}>0$, the majority fermions are spin-down
atoms in absence of SO coupling. The situation is similar as the first raw
of Fig. \ref{experiment}, with the only difference that atoms populate a
Fermi sea instead of occupying the lowest energy state only. In
one-dimension, the Fermi surface are simply two points whose momenta are
denoted by $k^{\pm}_\text{F}$ at positive and negative momenta,
respectively. Because the Raman process distorts the single particle
dispersion and breaks the spatial reflection symmetry, in general $k^{+}_%
\text{F}\neq k^{-}_\text{F}$. In Fig. \ref{Fermi} we show $\Delta k_\text{F}%
=k_\text{F}^{+}+k_\text{F}^{-}$ as a function of interaction modulation
strength. It shows that the direct Raman coupling and the interaction
assisted Raman coupling distort the fermion dispersion in an opposite way,
and therefore, $\Delta k_\text{F}$ changes from negative to positive when
interaction modulation increases. If $\delta<0 $ and $|\delta|>E_\text{F}$,
the majority fermions are spin-up atoms. The situation behaves as the second
raw of Fig. \ref{experiment}, and $\Delta k_\text{F}$ changes from negative
to positive when interaction modulation increases.

\textit{Method.} We consider modulating the interaction between two spin
component as $g\cos (2\omega t)\hat{n}_{\uparrow }(\mathbf{r})\hat{n}%
_{\downarrow }(\mathbf{r})$. For the situation we considered here, to very
good approximation, $\langle \hat{n}_{\uparrow }(\mathbf{r})+\hat{n}%
_{\downarrow }(\mathbf{r})\rangle $ is a constant. Therefore, it is
convenient to rewrite the interaction Hamiltonian as
\begin{equation}
-\frac{g}{4}\cos (2\omega t)(\hat{n}_{\uparrow }(\mathbf{r})-\hat{n}%
_{\downarrow }(\mathbf{r}))^{2}.
\end{equation}%
Here we leave the total density-density interaction for future consideration
as it does not enter the Raman processes considered here. We take the
mean-field approximation by defining the normalized magnetization as $M_{z}(%
\mathbf{r})=\langle \hat{n}_{\uparrow }(\mathbf{r})-\hat{n}_{\downarrow }(%
\mathbf{r})\rangle /n_{0}$, where $n_{0}=N/V$ is the average total density.
Then, this mean-field Hamiltonian can be written as
\begin{equation}
\hat{H}_{\text{MF}}=\hat{H}_{0}-\frac{g}{2}n_{0}M_{z}(\mathbf{r})\sigma
_{z}\cos (2\omega t),  \label{Hmf}
\end{equation}%
where $\hat{H}_{0}$ is given by Eq. (\ref{H0}). Note that here $M(\mathbf{r})
$ needs to be determined self-consistently.

We employ the Floquet approach to solve this mean-field Hamiltonian. We can
define a time evolution operator $\hat{U}(T)=\int_0^{T}dt\hat{H}_\text{MF}%
(t) $. For a given $M_z(\mathbf{r})$, we numerically diagonalizing the time
evolution operator as $\hat{U}(T)|\varphi_n\rangle=e^{-i\epsilon_n
T/\hbar}|\varphi_n\rangle$. Here $\epsilon_n$ is the quasi-energy, which is
restricted in the regime $-\pi\hbar/T<\epsilon_n<\pi\hbar/T$, and $%
|\varphi_n\rangle$ is the corresponding wave function. For bosons, we
consider that all atoms are condensed into the state with the lowest
quasi-energy. For fermions, we consider that all atoms fill a Fermi sea.
Then, we compute $M_z(\mathbf{r})$ either under this condensation wave
function or under the Fermi sea wave function, for bosons or fermions,
respectively. We iteratively solve the Floquet Hamiltonian until a
self-consistency is reached. Both Fig. \ref{Bose} and Fig. \ref{Fermi} are
obtained by this numerical method, as shown by the solid lines.

To see the physics more clearly, another way is to obtain the Floquet
effective Hamiltonian. Here we apply a unitary rotation $\hat{\mathcal{U}}%
_{2}$ to the Hamiltonian (\ref{Hmf}), and
\begin{equation}
\hat{\mathcal{U}}_{2}=e^{\frac{i}{\hbar }\int_{0}^{t}dt^{\prime }\left[
\frac{\hbar \omega }{2}\sigma _{z}-\frac{g}{2}n_{0}M_{z}\sigma _{z}\cos
(2\omega t^{\prime })\right] },
\end{equation}%
the Hamiltonian after rotation becomes
\begin{eqnarray}
\hat{H}(t) &=&-\frac{\hbar ^{2}\bigtriangledown ^{2}}{2m}+\delta \sigma
_{z}+\hbar \Omega \cos \left( 2k_{r}x-\omega t\right)   \notag \\
&&\times \left(
\begin{array}{cc}
0 & e^{i\left[ \omega t-\frac{\lambda M_{z}}{2}\sin \left( 2\omega t\right) %
\right] } \\
\text{h.c.} & 0%
\end{array}%
\right)   \label{rotation}
\end{eqnarray}%
where $\lambda =gn_{0}/(\hbar \omega )$. Here we have assumed that $M_{z}$
is a spatial independent constant, which has been well justified by the
numerical method described above. By only keeping the zeroth-order of Eq. (%
\ref{rotation}), we obtain a time-independent effective Hamiltonian as
\begin{eqnarray}
&&\hat{H}_{\text{eff}}=-\frac{\hbar ^{2}\bigtriangledown ^{2}}{2m}+\delta
\sigma _{z}+\frac{\hbar \Omega }{2}  \notag \\
&&\times \left(
\begin{array}{cc}
0 & \mathcal{J}_{0}\left( \frac{\lambda M_{z}}{2}\right) e^{i2k_{r}x}+%
\mathcal{J}_{1}\left( \frac{\lambda M_{z}}{2}\right) e^{-i2k_{r}x} \\
\text{h.c.} & 0%
\end{array}%
\right) .  \label{effective}
\end{eqnarray}%
Here $\mathcal{J}_{0}$ and $\mathcal{J}_{1}$ is the zeroth and the first
order Bessel functions. The numerical results based on Eq. (\ref{effective})
are shown by the dashed lines in Fig. \ref{Bose} and Fig. \ref{Fermi}, which
agree well with the solid lines.

The effective Hamiltonian Eq. (\ref{effective}) is a central result of this
work. It represents a faithful representation of the density-dependent SO
coupling proposed in this work, and illustrates clearly the competition of
two Raman processes. In the presence of both positive and negative momentum
transfers, the single particle dispersion in general should develop a band
structure. For small interaction modulation, when $\lambda \rightarrow 0$, $%
\mathcal{J}_{0}\rightarrow 1$ and $\mathcal{J}_{1}\rightarrow 0$. When we
ignore $\mathcal{J}_{1}$ term, this effective Hamiltonian recovers Eq. (\ref%
{direct_Raman}) upon a gauge transformation, with $\Omega $ replaced by $%
\Omega \mathcal{J}_{0}\left( \lambda M_{z}/2\right) $. As $\lambda $
increases, $\mathcal{J}_{1}$ increases and $\mathcal{J}_{0}$ decreases. For
large interaction modulation, when $\mathcal{J}_{1}$ term dominates, by only
keeping the $\mathcal{J}_{1}$ term, the effective Hamiltonian recovers Eq. (%
\ref{Interaction_Raman}), with $\tilde{\Omega}$ given by $\Omega \mathcal{J}%
_{1}\left( \lambda M_{z}/2\right) $.

\textit{Emergent $Z_{2}$ Symmetry.} The crossover between these two regimes
happens at $\mathcal{J}_{0}\left( \lambda M_{z}/2\right) \approx \mathcal{J}%
_{1}\left( \lambda M_{z}/2\right) $, and a special point is that $\mathcal{J}%
_{0}\left( \lambda M_{z}/2\right) =\mathcal{J}_{1}\left( \lambda
M_{z}/2\right) $. As we mentioned above, the presence of SO coupling
generally breaks the spatial reflection symmetry. However, this symmetry is
restored when $\mathcal{J}_{0}\left( \lambda M_{z}/2\right) =\mathcal{J}%
_{1}\left( \lambda M_{z}/2\right) $. As one can see, spatial reflection $%
x\rightarrow -x$, together with a $\pi $ spin rotation along $\hat{z}$, keep
the effective Hamiltonian Eq. (\ref{effective}) invariant.

This emergent $Z_{2}$ symmetry in the crossover regime has direct
experimental signatures. First, as illustrated in the middle column of Eq. (%
\ref{experiment}), the clouds of minority spin component appears at both
positive and negative momenta, which are of equal size. Secondly, in Fig. %
\ref{Bose} one can see that when $\lambda =3.04$, the condensation momentum $%
k_{\text{min}}=0$. Thirdly, in Fig. \ref{Fermi} one can see that when $%
\lambda =3.06$, $\Delta k_{\text{F}}=0$. Both are indicated by dashed lines
in Fig. \ref{Bose} and \ref{Fermi}. It is remarkable to note that although
Fig. \ref{Bose} and Fig. \ref{Fermi} consider two different systems with
different $\delta $, different dimensionality and different statistics,
these two values of $\lambda $ agree remarkably with each other, because
both are determined by the underlying spatial reflection symmetry. This
value is also consistent with the condition $\mathcal{J}_{0}\left( \lambda
M_{z}/2\right) \approx \mathcal{J}_{1}\left( \lambda M_{z}/2\right) $, which
gives $\lambda =2.87$ by taking $M_{z}\approx 1$. In fact, the
self-consistent $M_{z}$ is close to but smaller than unity when $\Omega <4E_{%
\text{r}}$, with $E_{\text{r}}=\hbar ^{2}k_{\text{r}}^{2}/(2m)$, and the
actual value of $\lambda $ for the emergent $Z_{2}$ symmetry is slightly
larger than $2.87$.

\textit{Outlook.} In the past decade, extensive studies have revealed rich
physics of ultracold atoms in the presence of a static SO coupling, whose
format and strength are both fixed. This work proposes a realistic proposal
to realize a density-dependent SO coupling, as given by Eq. (\ref{effective}%
). Since density of atoms is a dynamical field, this SO coupling has its
intrinsic dynamics. Here, as an important initial step to lay down the
basis, we only consider the mean-field theory, but more interesting effects
can certainly be found in future studies by including density fluctuations.
Novel effects can be found particularly in the regime either when density
fluctuations are strong, such as in interacting one-dimensional gases, or
when the system is sensitive to density, such as in the crossover regime.

\textit{Acknowledgment.} The project is supported by MOST under Grant No.
2016YFA0301600, Beijing Outstanding Young Scholar Program and NSFC Grant No.
11734010.


\begin{thebibliography}{99}
\bibitem{SOC_In_Solids@Kittel.1963} C. Kittel, \textit{Quantum Theory of
Solids}, (John Wiley and Sons Inc., 1963)

\bibitem{SOC_In_Solids@Winkler.2003} R. Winkler, \textit{Spin-orbit Coupling
Effects in Two-Dimensional Electron and Hole Systems}, (Springer-Verlag,
Berlin, Heidelberg 2003)

\bibitem{SOC_Rev@Zhai.2012} H. Zhai, Int. J. Mod. Phys. B \textbf{26},
1230001 (2012)

\bibitem{SOC_ReV@Spielman.2014} N. Goldman, G. Juzeli\={u}nas, P \"{O}hberg,
and I. B. Spielman, Rep. Prog. Phys. \textbf{77}, 126401 (2014)

\bibitem{SOC_Rev@Zhai.2015} H. Zhai, Rep. Prog. Phys. \textbf{78}, 026001
(2015)

\bibitem{SOC_BEC@Spielman.2011} Y. J. Lin, K. Jim\'{e}nez-Garc\'{\i}a, and
I. B. Spielman, Nature \textbf{471}, 83 (2011)

\bibitem{SOC_fermion@Zhang_Jin.2012} P. Wang, Z. Q. Yu, Z. Fu, J. Miao, L.
Huang, S. Chai, H. Zhai, and J. Zhang, Phys. Rev. Lett. \textbf{109}, 095301
(2012)

\bibitem{SOC_fermion@Zwierlein.2012} L. W. Cheuk, A. T. Sommer, Z.
Hadzibabic, T. Yefsah, W. S. Bakr, and M. W. Zwierlein, Phys. Rev. Lett.
\textbf{109}, 095302 (2012)

\bibitem{SOC_BEC@Galitski.2008} T. D. Stanescu, B. Anderson, and V.
Galitski, Phys. Rev. A \textbf{78} 023616 (2008)

\bibitem{SOC_BEC@Zhai.2010} C. Wang, C. Gao, C. M. Jian, and H Zhai, Phys.
Rev. Lett. \textbf{105}, 160403 (2010)

\bibitem{SOC_BEC@Jian_CM.2011} C. M. Jian, and H. Zhai, Phys. Rev. B \textbf{%
84}, 060508 (2011)

\bibitem{SOC_BEC@Ho_TL.2011} T.-L. Ho, and S. Zhang, Phys. Rev. Lett.
\textbf{107}, 150403 (2011)

\bibitem{SOC_BEC@Wu_CJ.2011} C. J. Wu, I. Mondragon-Shem, and X. F. Zhou,
Chin. Phys. Lett. \textbf{28}, 097102 (2011)

\bibitem{SOC_BEC@You_Li.2011} Z. F. Xu, R. L\"{u}, and L. You, Phys. Rev. A
\textbf{83}, 053602 (2011)

\bibitem{SOC_BEC@Santos.2011} S. Sinha, R. Nath, and L. Santos, Phys. Rev.
Lett. \textbf{107}, 270401 (2011)

\bibitem{SOC_BEC@Chen_Shuai.2012} J.-Y. Zhang, S.-C. Ji, Z. Chen, L. Zhang,
Z.-D. Du, B. Yan, G.-S. Pan, B. Zhao, Y. Deng, H. Zhai, S. Chen, J.-W. Pan,
Phys. Rev. Lett. \textbf{109}, 115301 (2012)

\bibitem{SOC_BEC@Baym_01.2012} T. Ozawa, and G. Baym, Phys. Rev. A \textbf{85%
}, 013612 (2012)

\bibitem{SOC_BEC@Baym_02.2012} T. Ozawa, and G. Baym, Phys. Rev. A \textbf{85%
}, 063623 (2012)

\bibitem{SOC_BEC@Li_Yun_01.2012} Y. Li, L. P. Pitaevskii, and S. Stringari,
Phys. Rev. Lett. \textbf{108}, 225301 (2012)

\bibitem{SOC_BEC@Martone.2012} G. L. Martone, Y. Li, L. P. Pitaevskii, and
S. Stringari, Phys. Rev. A \textbf{86}, 063621 (2012)

\bibitem{SOC_BEC@Zhang_Peng.2013} L. Zhang, J.-Y. Zhang, S.-C. Ji, Z. Du, H.
Zhai, Y. Deng, S. Chen, P. Zhang, and J.-W. Pan, Phys. Rev. A \textbf{87},
011601 (2013)

\bibitem{SOC_BEC@Clark.2013} R. M. Wilson, B. M. Anderson, and C. W. Clark,
Phys. Rev. Lett. \textbf{111}, 185303 (2013)

\bibitem{SOC_BEC@Demler.2013} S. Gopalakrishnan, I. Martin, and E. A.
Demler, Phys. Rev. Lett. \textbf{111}, 185304 (2013)

\bibitem{SOC_BEC@Li_Yun.2013} Y. Li, G. I. Martone, L. P. Pitaevskii, and S.
Stringari, Phys. Rev. Lett. \textbf{110}, 235302 (2013)

\bibitem{SOC_BEC@Zhou_Qi.2013} Q. Zhou, and X. Cui, Phys. Rev. Lett. \textbf{%
110}, 140407 (2013)

\bibitem{SOC_BEC_finite-T@Chen_Shuai.2014} S.-C. Ji, J.-Y. Zhang, L. Zhang,
Z.-D. Du1, W. Zheng, Y.-J. Deng, H. Zhai, S. Chen, and J.-W. Pan, Nat. Phys.
\textbf{10}, 1038 (2014)

\bibitem{SOC_BEC@Chen_Shuai.2015} S.-C. Ji, L. Zhang, X.-T. Xu, Z. Wu, Y.
Deng, S. Chen, J.-W. Pan, Phys. Rev. Lett. \textbf{114}, 105301 (2015)

\bibitem{SOC_BEC@Chen_Shuai.2016} Z. Wu, L. Zhang, W. Sun, X.-T. Xu, B.-Z.
Wang, S.-C. Ji, Y. Deng, S. Chen, X.-J. Liu, J.-W. Pan, Science \textbf{354}%
, 83 (2016)

\bibitem{SOC_FG@Shenoy.2011} J. P. Vyasanakere, S. Zhang, and V. B. Shenoy,
Phys. Rev. B \textbf{84}, 014512 (2011).

\bibitem{SOC_FG@Zhai.2011} Z. Q. Yu, and H. Zhai, Phys. Rev. Lett. \textbf{%
107}, 195305 (2011)

\bibitem{SOC_FG@Zhang.2011} M. Gong, S. Tewari, and C. Zhang, Phys. Rev.
Lett. \textbf{107}, 195303 (2011).

\bibitem{SOC_FG@Hu_Hui.2011} H. Hu, L. Jiang, X. J. Liu, and H. Pu, Phys.
Rev. Lett. \textbf{107}, 195304 (2011)

\bibitem{SOC_FG@Zhang_Peng.2012} P. Zhang, L. Zhang, and Y. Deng, Phys. Rev.
A \textbf{86}, 053608 (2012)

\bibitem{SOC_FG@Zhang_Peng.2013} L. Zhang, Y. Deng, and P. Zhang, Phys. Rev.
A \textbf{87}, 053626 (2013)

\bibitem{SOC_FG@Shenoy.2013} V. B. Shenoy, Phys. Rev. A \textbf{88}, 033609
(2013)

\bibitem{SOC_FG@Melo_01.2012} L. Han, and C. A. R. Sa' de Melo, Phys. Rev. A
\textbf{85}, 011606 (2012)

\bibitem{SOC_FG@Melo_02.2012} K. Seo, L. Han, and C. A. R. Sa' de Melo,
Phys. Rev. Lett. \textbf{109} 105303 (2012)

\bibitem{SOC_FG@Zhang_Jing.2014} Z. Fu, L. Huang, Z. Meng, P. Wang, L.
Zhang, S. Zhang, H. Zhai, P. Zhang, and J. Zhang, Nat. Phys. \textbf{10} 110
(2014)

\bibitem{SOC_FG@Zhang_Jing.2016} L. Huang, Z. Meng, P. Wang, P. Peng, S.-L.g
Zhang, L. Chen, D. Li, Q. Zhou, J. Zhang, Nat. Phys. \textbf{12}, 540 (2016)

\bibitem{Cavity_SOC@Yi_Su.2014} Y. Deng, J. Cheng, H. Jing, and S. Yi, Phys.
Rev. Lett. \textbf{112}, 143007 (2014)

\bibitem{Cavity_SOC@Yi_Wei.2015} J.-S. Pan, X.-J. Liu, W. Zhang, W. Yi, and
G.-C. Guo, Phys. Rev. Lett. \textbf{115}, 045303 (2015)

\bibitem{Cavity_SOC@Lev.2019} R. M. Kroeze, Y. Guo, B. L. Lev, Phys. Rev.
Lett. \textbf{123}, 160404 (2019)

\bibitem{Density_SOC@Chin.2018} L. W. Clark, B. M. Anderson, L. Feng, A.
Gaj, K. Levin, and C. Chin, Phys. Rev. Lett. \textbf{121}, 030402 (2018)

\bibitem{Density_SOC@Esslinger.2019} F. G\"{o}rg, K. Sandholzer, J.
Minguzzi, R. Desbuquois, M. Messer and T. Esslinger, Nat. Phys. \textbf{15},
1161 (2019)

\bibitem{Z2_LGT@Bloch.2019} C. Schweizer, F. Grusdt, M. Berngruber, L.
Barbiero, E. Demler, N. Goldman, I. Bloch, and M. Aidelsburger, Nat. Phys.
\textbf{15}, 1168 (2019)

\bibitem{footnote} There exists another situation that this $2\omega$ frequency term cannot be discarded. That is, there is an interaction energy $\sim 2\omega$ resonant with the $2\omega$ driving. This can also result in density dependent gauge field and this situation will be reported elsewhere.

\end{thebibliography}

\end{document}